\documentclass[prd,twocolumn,showpacs,amsmath,nofootinbib,amssymb,superscriptaddress]{revtex4}
\usepackage{graphicx}
\usepackage{dcolumn}
\usepackage{bm}
\makeatletter

\newcommand{\Rmnum}[1]{\expandafter\@slowromancap\romannumeral #1@}

\begin{document}

\title{Energy conditions bounds on  $f(T)$ gravity}

\author{Di Liu}
\affiliation{Center for Nonlinear Science and Department of Physics,
Ningbo University, Ningbo, Zhejiang, 315211 China}

\author{M.J. Rebou\c{c}as}
\affiliation{Centro Brasileiro de Pesquisas F\'{\i}sicas, \\
Rua Dr.\ Xavier Sigaud 150, 22290-180 Rio de Janeiro -- RJ, Brazil}

\date{\today}

\begin{abstract}
In the standard approach to cosmological modeling in the framework of general relativity,
the energy conditions play an important role in the understanding of several
properties of the  Universe, including singularity theorems, the current accelerating
expansion phase, and the possible existence of the so-called phantom fields.
Recently, the $f(T)$ gravity has been invoked as an alternative approach for
explaining the observed acceleration expansion of the Universe.
If gravity is described by a $f(T)$ theory instead of general relativity,
there are a number of issues that ought to be reexamined in the framework of
$f(T)$ theories. In this work, to proceed further with the current investigation
of the limits and potentialities of the  $f(T)$ gravity theories, we derive and
discuss the bounds imposed by the energy conditions on a general $f(T)$ functional
form. The null and strong energy conditions in the framework of  $f(T)$ gravity
are derived from first principles, namely the purely geometric Raychaudhuri
equation along with the requirement that gravity is attractive.
The weak and dominant energy conditions are then obtained in a direct
approach via an effective energy-momentum tensor for $f(T)$ gravity.
Although similar, the energy condition inequalities are different from those
of general relativity, but in the limit $f(T)=T$, the standard  forms
for the energy conditions in general relativity are recovered.
As a concrete application of the derived energy conditions to locally homogeneous
and isotropic $f(T)$ cosmology, we use the recent estimated values of the Hubble
and the deceleration parameters to set bounds from the weak energy condition
on the parameters of two specific families of $f(T)$ gravity theories.

\end{abstract}

\maketitle

\section{Introduction} \label{sect1}
A diverse set of cosmological observations coming from different sources, including the
supernovae-type Ia (SNe Ia)~\cite{sne}, the cosmic microwave background radiation
(CMBR)~\cite{cmbr}, and the large-scale structure (LSS)~\cite{lss} clearly indicate that
the Universe is currently expanding with an accelerating rate.
A number of alternative models and different frameworks have been proposed to account
for this observed late-time accelerated expansion of the Universe.
These approaches can be classified into two broad groups.
In the first, the framework of general relativity is kept unchanged
and an unknown form of matter sources, the so-called dark energy, is invoked.
In this regard, the simplest way to describe the accelerated expanding Universe
is by introducing a cosmological constant into the general relativity field equations.
Although this is entirely consistent with the available observational data,
it faces difficulties, including the microphysical origin and the order of magnitude
of the cosmological constant.
In the second group, modifications of Einstein's gravitation theory are assumed
as an alternative for describing the accelerated expansion.\footnote{An interesting member
of this group arises by assuming extra dimensions and by taking the Lagrangian of the theory
as function of the higher-dimensional Ricci scalar. This approach gives rise to the brane-world
cosmology~\cite{bw-refs}.}

Examples of the latter group include generalized theories of gravity based upon
modifications of the Einstein-Hilbert action by taking nonlinear functions $f(R)$
of the Ricci scalar $R$  or other curvature invariants (for reviews see Ref.~\cite{fr}).

An alternative modification of  general relativity, known as $f(T)$ gravity, has
been examined recently as a possible way of describing the current acceleration of
the Universe~\cite{Bengochea09,Linder10,Myrzakulov11}.
The origin of $f(T)$ gravity theory goes back to 1928 with Einstein's attempt
to unify gravity and electromagnetism through the introduction of a tetrad (vierbein)
field along with the concept of absolute parallelism or teleparallelism~\cite{Einstein}.
In the teleparallel gravity (TG) theories the dynamical object is not the metric
$g_{\mu \nu}$ but a set of tetrad fields $\mathbf{e^{}_{a}}(x^\mu)$, and rather than
the familiar torsionless Levi-Civita connection of general relativity, a Weitzenb\"{o}ck connection
(which has no curvature but only torsion) is used to define the covariant derivative.
The gravitational field equation of TG is then described in
terms of the torsion instead of the curvature~\cite{FNGtn1,FNGtn2,FNGtn3}.
In formal analogy  with the $f(R)$,  the $f(T)$ gravity theory
was suggested by extending the Lagrangian of teleparallel gravity to a
function $f(T)$ of a torsion scalar $T$~\cite{Bengochea09,Linder10}. In comparison
with $f(R)$ gravity in the metric formalism, whose field equations are of the fourth order,
$f(T)$ gravity has the advantage that the dynamics are governed by  second-order
field equations.

The fact that $f(T)$ theories can potentially  be used to explain the observed accelerating
expansion along with the relative simplicity of their field equations has given birth to a
number of papers on these gravity theories, in which several features of $f(T)$ gravity have been
discussed, including observational cosmological constraints~\cite{Bengochea-2011,Wei-Ma-Qi-2011,Wu-Yu-2010},
solar system constraints~\cite{Iorio-Saridakis-2012}, cosmological perturbations~\cite{Dent-Duta-Saridakis-2011,Zheng-Huang-2011,Chen-dent-Dutta-2011}, dynamical behavior~\cite{Wu-Yu-b-2010},
spherically symmetric solutions~\cite{Wang-2011},
the existence of relativistic stars~\cite{Stars-in-f(T)}, the possibility of quantum divide
crossing~\cite{Wu-Yu-2011}, cosmographic constraints~\cite{Cosmography-2011},
and the lack of local Lorentz invariance~\cite{Li-Sotiriou-Barrow-2011,Miao-Li-Miao-2011,Li-Miao-Miao-2011}
which may give rise to undesirable outcomes from $f(T)$ gravity~\cite{Zheng2011,Lilarg},
although suitable tetrad fields can be chosen~\cite{Tamanini-Bohmer-2012}.
For some further references on several aspects of $f(T)$ gravity theories we refer the
readers to Ref.~\cite{FT}.

In the framework of general relativity the so-called \emph{energy conditions}
have been used to derive remarkable results in a number of contexts. For
example, the famous Hawking-Penrose singularity theorems invoke the strong
energy condition (SEC)~\cite{Hawking-Ellis}, whose violation allows for the
observed accelerating expansion, and the proof of the second law of
black hole thermodynamics  requires
null energy conditions (NEC)~\cite{Visser,Wald}.

On macroscopic scales relevant for cosmology, the confrontation of the energy
conditions predictions with observational data is another important issue that
has been considered in a number of recent articles.
In this regard, since the pioneering works by Visser~\cite{M_Visser1997},
a number of articles have been published concerning this confrontation by using
model-independent energy-conditions bounds on the cosmological
observable quantities, such as the distance modulus, lookback time,
and deceleration and curvature parameters~\cite{EC-1,EC-2,EC-3,EC-4,EC-5,EC-6,EC-7,EC-8}.

Owing to their role in several important issues in general relativity and cosmology,
the energy conditions have also been investigated in several frameworks of
modified gravity theories, including $f(R)$ gravity~\cite{EC-fR-grav,Kung},
gravity with nonminimal coupling between curvature and matter~\cite{Bertolami-Sequeira09},
Gauss-Bonnet gravity~\cite{EC-Gauss-Bonnet}, modified $f(G)$ gravity~\cite{EC_fG-grav},
and Brans-Dicke theories~\cite{EC-Brans-Dicke}
(see also the related Refs.~\cite{EC_fG-NM-copling,EC-f(RT)}).

In this article, to proceed further with these investigations on the potentialities,
difficulties, and limitations of $f(T)$ gravity theories, we derive the energy conditions
for the general functional form of $f(T)$ and discuss some concrete examples of these
bounds by using observational constraints on the Hubble and the deceleration parameters.
The null and strong energy conditions (NEC and SEC) are derived  in the framework of
$f(T)$ from first principles, i.e., from the purely geometric Raychaudhuri equation
along with the requirement that gravity is attractive.
We find that the NEC and the SEC in general $f(T)$ gravity, although similar, are
different from those of Einstein's gravity and $f(R)$ gravity, but in the limiting
case $f(T)= T$, the standard general relativity forms for theses energy conditions
are recovered.
The resulting inequalities for the SEC and NEC in the $f(T)$ gravity framework are then
compared with what would be obtained by translating these energy conditions in terms of
an effective energy-momentum tensor for $f(T)$ gravity.There emerges from this comparison a natural
formulation for the weak and dominant energy conditions (WEC and DEC) in the context
of $f(T)$ gravity, which also reduce to the standard GR forms for these conditions
in the limit $f(T) = T$. As a concrete application of the energy conditions for spatially homogeneous and
isotropic $f(T)$ cosmology, we use recent estimated values of the Hubble
and the deceleration parameters to set bounds from the WEC
on the parameters of two specific families of $f(T)$ gravity theories.

Our paper is organized as follows. In Sec.~\ref{sec2}, we give a brief review on
the $f(T)$ theories and derive the field equations. In Sec.~\ref{sec3}, using
the purely geometric Raychaudhuri equations for timelike and null congruences
of curves, we derive the SEC and NEC from first principles, and the WEC and DEC
through an effective energy-momentum tensor. In Sec.~\ref{sec4} we use the constraints
on present-day values of cosmographic parameters to set constraints on
exponential as well as on the Born-Infeld $f(T)$ gravity from the WEC.
Finally, conclusions and final remarks are presented in Sec.~\ref{sec5}.
\section{$\mathbf{f(T)}$ gravity theory} \label{sec2}

In this section, we  briefly introduce the teleparallel gravity and its
generalization known as $f(T)$ gravity. We begin by recalling that
the dynamical variables in teleparallel gravity are the vierbein or tetrad fields,
$\mathbf{e_a}(x^\mu)$, which is a set of four ($a= 0,\cdots,3$) vectors defining
a local orthonormal frame at every point $x^\mu$ of the spacetime manifold.
The tetrad vectors field $\mathbf{e_a}(x^\mu)$ are vectors in the tangent space and
can be expressed in terms of a coordinate basis as $\mathbf{e_a}(x^\mu)=\,e^\mu_a\partial_\mu$.
The spacetime  metric tensor and the tetrads are related by%
\footnote{Throughout this paper we use Greek letters to denote spacetime coordinate indices,
which are lowered and raised, respectively,  with $g_{\mu \nu }$ and $g^{\mu \nu }$, and vary from
$0$ to $3$,  whereas
firsts alphabetic  latin lower case letters ($a$ and $b$) are tetrad indices,  which are lowered and raised
with the Minkowski tensor $\eta_{ab} = \text{diag}\,(1,-1,-1,-1)$ and $ \eta^{ab}$, respectively.
We denote the spatial components ($1,  2, 3$) by using the middle alphabetic latin lower case
letters $i$ and $j$.}

\begin{equation} \label{g-metric}
g_{\mu \nu }=e_{\mu }^{a}\,e_{\nu }^{b}\,\eta _{ab}
\end{equation}
where $\eta _{ab}=\text{diag}\,(1,-1,-1,-1)$ is the Minkowski metric of the tangent space at $x^\mu$.
It follows that the relation between frame components,  $e_{a}^{\mu }$, and coframe components,
$e_{\mu }^{a}$, are given by
\begin{equation}\label{tetradralation}
e_{a}^{\mu }\,e_{\nu}^{a}=\delta _{\nu }^{\mu } \qquad \text{and} \qquad
e_{a}^{\mu }\,e_{\mu }^{b}=\delta _{a}^{b}\;.
\end{equation}

In general relativity one uses the Levi-Civita connection
\begin{equation}
\overset{\circ }{\Gamma }{}_{\;\;\mu \nu }^{\rho } =
\frac{1}{2}g^{\rho \sigma }\left(
\partial _{\nu} g_{\sigma \mu}+\partial _{\mu}g_{\sigma \nu}-\partial _{\sigma}g_{\mu \nu}\right)\;,
\end{equation}%
which  leads to nonzero spacetime curvature but zero torsion.

In teleparallel gravity, instead of the Levi-Civita connection, one uses
the Weitzenb\"{o}ck connection which is given by
\begin{equation}\label{connection}
\widetilde{{\Gamma }}_{\;\mu \nu }^{\lambda }=e_{a}^{\lambda }\,\partial _{\nu }\,
e_{\mu}^{a} = -e_{\mu }^{a}\,\partial _{\nu }\,e_{a}^{\lambda }\;.
\end{equation}
An immediate consequence of this definition is that the covariant derivative, $D_{\mu }$, of
the tetrad fields
\begin{equation}
D_{\mu }e_{\nu }^{a} \equiv \partial _{\mu }e_{\nu }^{a}-
\widetilde{\Gamma} _{\;\nu \mu}^{\lambda }e_{\lambda }^{a}=0\;,
\end{equation}
vanishes identically. This equation leads to a zero curvature but nonzero torsion.

To clarify the interrelations between Weitzenb\"{o}ck and Levi-Civita connections,
one needs to introduce  the torsion and contorsion tensors, which are given,
respectively, by
\begin{equation}\label{T}
T^{\rho }_{\;\;\mu \nu } \equiv \widetilde{\Gamma }_{\;\nu \mu }^{\rho }
-\widetilde{\Gamma }_{\;\mu \nu}^{\rho } =
e_{a}^{\rho }(\partial _{\mu }e_{\nu }^{a}-\partial _{\nu }e_{\mu}^{a})\;,
\end{equation}
\begin{equation}\label{K}
K_{\;\;\mu \nu }^{\rho } \equiv \widetilde{\Gamma} _{\;\mu \nu }^{\rho }
-\overset{\circ}{\Gamma }{}_{\;\mu \nu }^{\rho}=\frac{1}{2}(T_{\mu }{}^{\rho }{}_{\nu }
+ T_{\nu}{}^{\rho }{}_{\mu }-T_{\;\;\mu \nu }^{\rho })\;,
\end{equation}
where above $\overset{\circ }{\Gamma }{}_{\;\;\mu \nu }^{\rho }$ is the
Levi-Civita connection. 

Now, if one further defines the so-called super-potential
\begin{equation}\label{S}
S_{\sigma }^{\;\;\mu \nu }\equiv K_{\;\;\;\;\sigma }^{\mu \nu
}+\delta _{\sigma }^{\mu }T_{\;\;\;\;\;\xi }^{\xi \nu }-\delta _{\sigma
}^{\nu }T_{\;\;\;\;\;\xi }^{\xi \mu }\;,
\end{equation}
one obtains the torsion scalar
\begin{equation}\label{scalarT}
T\equiv \frac{1}{2}S_{\sigma }^{\;\;\mu \nu }T_{\;\;\mu \nu }^{\sigma }=
\frac{1}{4}T^{\xi \mu \nu }T_{\xi \mu \nu }+\frac{1}{2}T^{\xi \mu
\nu }T_{\nu \mu \xi }-T_{\xi \mu }^{\;\;\;\;\xi }
T_{\;\;\;\;\;\nu }^{\nu \mu}\,,
\end{equation}
which is used as the Lagrangian density in formulation of the teleparallel
gravity theory, which is given by

\begin{equation}
\mathcal{L}_{T}  = \frac{e\,T}{2\,\kappa^{2}}\;,
\end{equation}
where $e=\det (e_\mu^{a})=\sqrt{-g}$, $\kappa^2=8\pi G$, and $G$ is the
gravitational constant.
Now, by taking an arbitrary function $f$ of the torsion scalar $T$, one obtains
the Lagrangian density of $f(T)$ gravity theory, that is
\begin{equation}\label{f1}
\mathcal{L}_{T} \,\longrightarrow \,\mathcal{L}_{f(T)} = \frac{e\,f(T)}{2\,\kappa^{2}}.
\end{equation}
Now, by adding a matter Lagrangian density $\mathcal{L}_M$ to Eq.~(\ref{f1}) and
varying the resultant action with respect to the vierbein, one obtains the
following field equation for $f(T)$ gravity:
\begin{eqnarray}\label{lagran1}
&&\partial_\xi(ee^\rho_a S^{\;\;\sigma\xi}_\rho f_T)-ee^\lambda_a S^{\rho\xi\sigma} T_{\rho\xi\lambda}f_T+\frac{1}{2}ee^\sigma_af(T) \nonumber \\
&&=[\partial_\xi(ee^\rho_a
S^{\;\;\sigma\xi}_\rho)-ee^\lambda_a S^{\rho\xi\sigma} T_{\rho\xi\lambda}]f_T
+ee^\rho_a(\partial_\xi T)S^{\;\;\sigma\xi}_\rho f_{TT} \nonumber \\
&&+\frac{1}{2}ee^\sigma_af(T) = e\,\Theta^\sigma_{\;\;a}\;,
\end{eqnarray}
where $f_T= df(T)/dT$, $f_{TT}= d^2f(T)/dT^2$, and $\Theta^\sigma_{\;\;a}$ is the
energy-momentum tensor of the matter fields. Here and in what follows we have chosen
units such that $\kappa^2=c=1$.

To bring the field equations (\ref{lagran1}) to a form suitable for our purpose
in the next section. To this end, we first note that if one multiply \textbf{$e^{-1}g_{\mu\sigma}e^a_\nu$},
both sides of (\ref{lagran1}), the resultant equation is such that the coefficient of that the term
$f_{T}$ takes the form
\begin{eqnarray}\label{nablaS}
&&e^a_\nu e^{-1}\partial_\xi(ee^\rho_a
S^{\;\;\sigma\xi}_\rho)-S^{\rho\xi\sigma}T_{\rho\xi\nu} \nonumber \\
&&=\partial_\xi S^{\;\;\sigma\xi}_{\nu}-\widetilde{\Gamma }_{\;\nu \xi }^{\rho }S^{\;\;\sigma\xi}_{\rho}
+\overset{\circ}{\Gamma }{}_{\;\tau\xi}^{\tau}S^{\;\;\sigma\xi}_{\nu}
-S^{\rho\xi\sigma}T_{\rho\xi\nu} \nonumber \\
&&=-\nabla^\xi S_{\nu\xi}^{\;\;\;\;\sigma}-S^{\xi\rho\sigma}K_{\rho\xi\nu}\;,
\end{eqnarray}
where the relation
\begin{equation} \label{relation1}
K^{(\mu\nu)\sigma}=T^{\mu(\nu\sigma)}=S^{\mu(\nu\sigma)}=0
\end{equation}
has been used.

On the other hand, from the relation between and Weitzenb\"ock connection and the
Levi-Civita connection given by Eq.(\ref{K}), one can write the Riemann tensor for the
Levi-Civita connection in the form  
\begin{eqnarray}\label{tensorR}
R^\rho_{\;\;\mu\lambda\nu}\!\!\!\!\!\!&&=\partial_{\lambda}\overset{\circ}{\Gamma }{}_{\;\mu\nu}^{\rho}
-\partial_{\nu}\overset{\circ}{\Gamma }{}_{\;\mu\lambda}^{\rho}
+\overset{\circ}{\Gamma }{}_{\;\sigma\lambda}^{\rho}\overset{\circ}{\Gamma }{}_{\;\mu\nu}^{\sigma}
-\overset{\circ}{\Gamma }{}_{\;\sigma\nu}^{\rho}\overset{\circ}{\Gamma }{}_{\;\mu\lambda}^{\sigma}\\ \nonumber
&&=\nabla_\nu K^\rho_{\;\;\mu\lambda}-\nabla_\lambda K^\rho_{\;\;\mu\nu}
+K^\rho_{\;\;\sigma\nu}K^\sigma_{\;\;\mu\lambda}-K^\rho_{\;\;\sigma\lambda}K^\sigma_{\;\;\mu\nu}\;,
\end{eqnarray}
whose associated Ricci tensor can then be written as
\begin{equation}
R_{\mu\nu}=\nabla_\nu K^\rho_{\;\;\mu\rho}-\nabla_\rho K^\rho_{\;\;\mu\nu}
+K^\rho_{\;\;\sigma\nu}K^\sigma_{\;\;\mu\rho}
-K^\rho_{\;\;\sigma\rho}K^\sigma_{\;\;\mu\nu}\;.
\end{equation}
Now, by using $K^\rho_{\;\;\mu\nu}$ given by Eq.~(\ref{S}) along with the
relations~(\ref{relation1}) and considering that $S^\mu_{\;\;\rho\mu}=
2K^\mu_{\;\;\;\rho\mu}=-2T^\mu_{\;\;\;\rho\mu}$ one has
~\cite{Li-Sotiriou-Barrow-2011,Sotiriou-Li-Barrow2011b,Lilarg}
\begin{eqnarray}
&&R_{\mu\nu}=-\nabla^\rho S_{\nu\rho\mu}-g_{\mu\nu}\nabla^\rho T^\sigma_{\;\;\;\rho\sigma}
-S^{\rho\sigma}_{\;\;\;\;\;\mu}K_{\sigma\rho\nu}\;, \nonumber \\
&&R=-T-2\nabla^\mu T^\nu_{\;\;\;\mu\nu}\;,
\end{eqnarray}
and thus obtain
\begin{equation}\label{eqdivs}
G_{\mu\nu}-\frac{1}{2}\,g_{\mu\nu}\,T
=-\nabla^\rho S_{\nu\rho\mu}-S^{\sigma\rho}_{\;\;\;\;\mu}K_{\rho\sigma\nu}\;,
\end{equation}
where $G_{\mu\nu}=R_{\mu\nu}-(1/2)\,g_{\mu\nu}\,R$ is the Einstein tensor.

Finally, combining Eq.~(\ref{nablaS}) and Eq.~(\ref{eqdivs}), the field equations for
$f(T)$ gravity Eq.~(\ref{lagran1}) can be rewritten in the form
\begin{equation}\label{motion1}
A_{\mu\nu}f_T+B_{\mu\nu}f_{TT}+\frac{1}{2}g_{\mu\nu} f(T) =\Theta_{\mu\nu}\;,
\end{equation}
where
\begin{eqnarray}\label{motion1add}
&&A_{\mu \nu }=g_{\sigma\mu}e^a_\nu[e^{-1}\partial_\xi(ee^\rho_a
S^{\;\;\sigma\xi}_\rho)-e^\lambda_a S^{\rho\xi\sigma} T_{\rho\xi\lambda}]\\ \nonumber
&&\qquad=-\nabla^\sigma S_{\nu\sigma\mu }-S_{\;\;\;\;\mu }^{\rho\lambda }K_{\lambda \rho \nu }
=G_{\mu \nu }-\frac{1}{2}g_{\mu \nu }T, \; \\ \nonumber
&&B_{\mu\nu}=S^{\;\;\;\;\sigma}_{\nu\mu}\nabla_\sigma T\;.
\end{eqnarray}

To close this section, we note that since $A_\mu^{\;\;\mu}=-(R+2T)$, the trace of
Eq.~(\ref{motion1}), which can be used as an independent relation to simplify the field
equation,  can be expressed as
\begin{eqnarray}\label{trace}
-(R+2T)f_T + B f_{TT} + 2f(T)=\Theta\;,
\end{eqnarray}
where $B=B_\mu^{\;\;\mu}$ and $\Theta=\Theta_\mu^{\;\;\mu}$.

\section{Energy Conditions} \label{sec3}

\subsection{Strong and null energy conditions}

The ultimate origin of strong and null energy conditions is the Raychaudhuri
equation together with the requirement that gravity is attractive. The Raychaudhuri
equation gives temporal variation of the expansion $\theta$ of congruence of geodesics
(for a review article see Ref.~\cite{Kar-Dadhich}).
For a congruence of timelike geodesics whose tangent vector field is $u^{\mu}$
Raychaudhuri equation  reads
\begin{equation}  \label{Raych-time}
\frac{d\theta}{d\tau}= - \frac{1}{3}\,\theta^2 -
\sigma_{\mu\nu}\sigma^{\mu\nu} + \omega_{\mu\nu}\omega^{\mu\nu}
- R_{\mu\nu}u^{\mu}u^{\nu} \;,
\end{equation}
where   $\theta\,$, $\sigma^{\mu\nu}$ and $\omega_{\mu\nu}$ are, respectively, the  expansion, shear,
and rotation associated with the congruence defined by the vector field $u^{\mu}$, and $R_{\mu\nu}$
is the Ricci tensor.

The evolution equation for the expansion of a congruence of null geodesics
defined by a null vector field $k^\mu$ has a similar form as the Raychaudhuri
equation (\ref{Raych-time}), but with a factor $1/2$ rather than $1/3$, and
$-R_{\mu\nu}k^{\mu}k^{\nu}$ instead of $-R_{\mu\nu}u^{\mu}u^{\nu}$
as the last term (see Ref.~\cite{Carroll} for more details). Thus,
its reads
\begin{equation}\label{Raych-null}
\frac{d\theta}{d\tau}=-\frac{1}{2}\,\theta^2-
\sigma_{\mu\nu}\sigma^{\mu\nu}+\omega_{\mu\nu}\omega^{\mu\nu}
- R_{\mu\nu}k^\mu k^\nu \;,
\end{equation}
where the kinematical quantities $\theta\,$, $\sigma^{\mu\nu}$ and
$\omega_{\mu\nu}$ are now clearly associated with the congruence of
null geodesics.

An important point to be emphasized is that Raychaudhuri Eqs.~(\ref{Raych-time})
and~(\ref{Raych-null}) are purely geometric statements, and as such they make
no reference to any theory of gravitation.

Now, since the shear is a "spatial" tensor, i.e., $\sigma^2 \equiv \sigma_{\mu\nu}
\sigma^{\mu\nu}\geq 0$, from Eqs.~(\ref{Raych-time}) and~(\ref{Raych-null}), one
has that for any hypersurface of orthogonal congruences ($\omega_{\mu\nu}=0$),
the conditions for gravity to remain attractive ($d\theta / d\tau < 0$) are given by
\begin{eqnarray}
R_{\mu\nu}u^\mu u^\nu\geq 0 \;, \label{SEC} \\
R_{\mu\nu}k^\mu k^\nu \geq 0 \;. \label{NEC}
\end{eqnarray}
Thus, as long as one can use the field equations of any given gravity theory
to relate $R_{\mu \nu}$ to the energy-momentum tensor $T_{\mu \nu}$,
the above Raychaudhuri Eqs.~(\ref{Raych-time}) and~(\ref{Raych-null}),
along with the requirement that gravity is attractive, lead to
Eqs.~(\ref{SEC}) and~(\ref{NEC}), which can be employed to restrict the
energy-momentum tensors in the framework of the gravity theory one is
concerned with.

Equations~(\ref{SEC}) and~(\ref{NEC}) are ultimately the SEC and DEC
stated in a coordinate-invariant way for an unfixed
geometrical theory of gravitation. Hence, for example, in the framework of
general relativity, they take, respectively, the forms%
\footnote{Clearly, here $T$ is not the torsion scalar, but the trace of
the energy momentum tensor $T=T^\mu_\mu$.}
\begin{eqnarray}
R_{\mu\nu}\, u^{\mu}u^{\nu}&=&\left( T_{\mu\nu} - \frac{T}{2}
g_{\mu\nu} \right)\,u^{\mu}u^{\nu}\geq 0 \,,  \label{StrongEC}
\end{eqnarray}
and
\begin{eqnarray} \label{NullEC}
R_{\mu\nu}k^{\mu}k^{\nu} = T_{\mu\nu}\, k^{\mu}k^{\nu}\geq 0\,,
\end{eqnarray}
which, for example,  for  a perfect fluid of density $\rho$ and pressure $p\,$, i.e.,
for $T_{\mu\nu}= (\rho + p)\,u_{\mu}u_{\nu} - p\,g_{\mu\nu},$ reduce to the well-known
forms of the SEC and NEC in general relativity
\begin{equation} \label{SEC-GR}
\rho + 3p \geq 0  \qquad \text{and} \qquad
\rho + p  \geq  0 \;.
\end{equation}

\subsection{Energy conditions in $f(T)$ gravity} \label{ssec3}

According to the previous section the Raychaudhuri equations together
with the attractive character of the gravitational interaction give rise to
Eqs.~(\ref{SEC}) and~(\ref{NEC}), which hold for any geometrical
theory of gravitation. In what follows, we maintain this approach to
derive the SEC and NEC in the $f(T)$ gravity context. To this end,
we first rewrite the $f(T)$ field equation~(\ref{motion1}) in the
form
\begin{eqnarray}\label{einst}
G_{\mu\nu}=\frac{1}{f_T}[\,\Theta_{\mu\nu}+\frac{1}{2}(Tf_T-f)g_{\mu\nu}
- B_{\mu\nu}f_{TT}\,]\,.
\end{eqnarray}
Here, $\Theta_{\mu\nu}$ and $\Theta$ denote, respectively,
the energy momentum tensor and its trace.

{}From Eq.~(\ref{einst}) and by taking into account the trace
equation~(\ref{trace}), we have
\begin{eqnarray}
R_{\mu\nu}=\mathcal{T}_{\mu\nu}-\frac{1}{2}\,g_{\mu\nu}\,\mathcal{T}\;,
\end{eqnarray}
where \vspace{-3mm}
\begin{eqnarray}\label{TTT}
\mathcal{T}_{\mu\nu} &=&\frac{1}{f_T}(\Theta_{\mu\nu}-f_{TT}B_{\mu\nu})\;, \\
\mathcal{T}&=&\frac{1}{f_T}(\Theta + T f_T - f- B f_{TT}).
\end{eqnarray} \vspace{0.1mm}

Now, for the homogeneous and isotropic Friedmann-Lema\^{\i}tre-Robertson-Walker
(FLRW) metric with scale factor $a(t)$, i.e., $g_{\mu\nu}=diag(1,-a^2,-a^2,-a^2)$, from
Eqs.~(\ref{T}) through (\ref{scalarT}) along with Eq.~(\ref{motion1add}), we have
\begin{eqnarray}
T=-6H^2\,,
\end{eqnarray}
\begin{eqnarray}\label{Amu}
A_{00}=6H^2\,,  & \qquad  A_{ij}=-2a^2(3H^2+\dot{H})\,\delta_{ij}\,,
\end{eqnarray}
\begin{eqnarray}
B_{ij}=24a^2H^2\dot{H}\,\delta_{ij}\,,   \qquad B=-72H^2\dot{H}\,,
\end{eqnarray}
where a dot denotes derivative with respect to time, $H=\dot{a}/a$ is the Hubble
parameter, and the simplest and suitable tetrad basis was used~\cite{Tamanini-Bohmer-2012}.

Now, for a perfect fluid of density $\rho$ and pressure $p$, namely for
\begin{eqnarray}\label{emt}
\Theta_{\mu\nu}=(\rho+p)u_\mu u_\nu-p\,g_{\mu\nu} \;\; \text{with} \;\; u_\mu=(1,0,0,0)\,,
\end{eqnarray}
taking $k_\mu=(1, a, 0, 0)$, we obtain the $\mathcal{T}_{\mu\nu}\,$ and its trace
$\mathcal{T}$, namely
\begin{eqnarray}
\mathcal{T}_{00}=\frac{1}{f_T}\,\rho\,, \quad \; \mathcal{T}_{ij}
=\frac{a^2}{f_T}(p-24H^2\dot{H}f_{TT})\,\delta_{ij}\;,
\end{eqnarray}
and
\begin{equation}
\mathcal{T}=\frac{1}{f_T}(\rho-3p+Tf_T-f+72H^2\dot{H}f_{TT})\;.
\end{equation}

Thus, from equations~(\ref{SEC}) and (\ref{NEC}) for a general $f(T)$ gravity,
the strong energy condition (SEC) and the null energy condition (NEC) can be,
respectively, written as
\begin{eqnarray} \label{sec1}
\!\!\!\!\!\!\! \mbox{\bf SEC}:\quad \frac{1}{2f_T}(\rho+3p+f-Tf_T-72H^2\dot{H}f_{TT})\geq 0\,,
\end{eqnarray}
and
\begin{eqnarray}\label{nec1}
\!\!\!\!\!\!\!\!\!\!\!\mbox{\bf NEC}: \quad \frac{1}{f_T}(\rho+p-24H^2\dot{H}f_{TT})\geq 0 \,.
\end{eqnarray}
We note that the well-known forms for the SEC ($\rho + 3p  \geq 0$) and NEC ($\rho + p  \geq 0$)
in the framework of general relativity can be recovered as a particular case of the
above SEC and DEC  in the context of $f(T)$ gravity for the special case $f(T)=T$,
as one would expect.

To derive the weak and dominant energy conditions (WEC and DEC)
in $f(T)$ gravity, it is important to realize that  the above SEC and NEC inequalities
[Eqs.(\ref{sec1}) and (\ref{nec1})] can also be recast as an extension of
the SEC and NEC conditions in the context of general relativity by defining
suitably an effective energy-momentum tensor in the context of $f(T)$ gravity.
In fact, in $f(T)$ gravity theories one can define an effective energy-momentum
tensor as
\footnote{A comparison with the effective energy-momentum tensor of Ref.~\cite{Lilarg}
makes clear that the one used in the present work includes the whole matter.}
\begin{eqnarray}
\Theta^{eff}_{\mu\nu}=\frac{1}{f_T}[\Theta_{\mu\nu}+\frac{1}{2}(Tf_T-f)g_{\mu\nu}
-f_{TT}B_{\mu\nu}]\;,
\end{eqnarray}
from which one defines the effective energy density and the effective pressure
in the FLRW by
\begin{eqnarray}\label{effrho}
\rho^{eff} = -g^{00}\Theta^{eff}_{00}=\frac{1}{f_T}[\rho+\frac{1}{2}(Tf_T-f)]\;,
\end{eqnarray}
\begin{eqnarray}\label{effp}
p^{eff}  &=& \frac{1}{3}g^{ij}\Theta^{eff}_{ij} \nonumber \\
         &=&\frac{1}{f_T} [p-\frac{1}{2}(Tf_T-f)- 24H^2\dot{H}f_{TT}]\;,
\end{eqnarray}
which in turn make apparent that the SEC and NEC given by Eqs.~(\ref{sec1})
and~(\ref{nec1}) can be obtained from  the corresponding general relativity
expressions [Eq.~(\ref{SEC-GR})] by using the above effective matter
components. Thus, using the effective energy-momentum tensor approach,
the  weak energy condition (WEC) in $f(T)$ gravity ($\,\rho^{eff}\geq 0\,$)
reduce to
\begin{eqnarray}\label{wec1}
\!\!\!\!\!\!\!\!\!\! \mbox{\bf WEC:} \quad \frac{1}{f_T}[\rho+\frac{1}{2}(Tf_T-f)]\geq 0\;.
\end{eqnarray}
Similarly, the dominant energy condition (DEC) in $f(T)$ gravity ($\,\rho^{eff}\geq|\,p\,|\,$)
can be written in the form
\begin{eqnarray}\label{dec1}
\!\!\!\!\!\!\! \mbox{\bf DEC:} \quad \frac{1}{f_T}[\rho-p+(Tf_T-f)+24H^2\dot{H}f_{TT}]\geq 0\;.
\end{eqnarray}

\section{Constraining $\mathbf{f(T)}$ gravity theories}\label{sec4}

The energy conditions~(\ref{sec1}), (\ref{nec1}),  (\ref{wec1}), and~(\ref{dec1})
can  be used to place bounds on a given $f(T)$ in the context of FLRW models. To investigate
such bounds, we first note that to ensure the positivity of the effective Newton gravity
constant, one has $f_T>0$~\cite{Zheng2011}.
Thus, after some algebra, in terms of present-day values for the cosmological parameters,
the energy conditions ~(\ref{sec1}), (\ref{nec1}),  (\ref{wec1}), and~(\ref{dec1}) can be,
respectively, rewritten as
\begin{eqnarray}\label{sec-2}
&\mbox{\bf SEC:} & \nonumber \\
\rho_0&\!\!\!+3p_0+f_0+6H_0^2f_{T_0}+72(1+q_0)H_0^4f_{T_0T_0}\geq 0\,; \\
\nonumber \\
&\mbox{\bf NEC:} & \nonumber \\
&\rho_0+p_0+24(1+q_0)H_0^4f_{T_0T_0}\geq0\,;\label{nec2} \\
\nonumber \\
&\mbox{\bf WEC:} & \nonumber \\
& 2\rho_0-f_0-6H_0^2f_{T_0}\geq0\,; \label{wec2} \\
\nonumber \\
&\mbox{\bf DEC:} & \nonumber \\
 \rho_0&\!\!\!\!\!-p_0-f_0-6[H_0^2f_{T_0}+4(1+q_0)H_0^4f_{T_0T_0}]\geq0\,, \label{dec2}
\end{eqnarray}
where $q = - (\ddot{a}/a)\,H^{-2}$ is the deceleration parameter, and a subscript $0$
indicates the present-day value of the corresponding parameter.

To make concrete applications of the above conditions to set bounds on $f(T)$, we first note
that apart from the WEC [Eq.~(\ref{wec2})], all the above conditions depend on the current
value of the pressure $p_0$.
Therefore, for simplicity
in what follows we shall focus on the observational WEC  constraints on $f(T)$ gravity.
Furthermore, we will also take the best fit value $H_0=0.718$
as determined by Cappozzielo {\it et al.}~\cite{Cosmography-2011}.

\subsection{Exponential $f(T)$ gravity} \label{subsec41}

As a first concrete example, we shall examine the WEC bounds on the parameter $\beta$
of the following exponential family of  $f(T)$ gravity
theories~\cite{Linder10,Wei-Ma-Qi-2011,kazu2011}:
\begin{eqnarray} \label{exp-grav}
f(T)=T+\alpha T(1-e^{\beta T_0/T})\;
\end{eqnarray}
with
\begin{eqnarray}
\alpha=-\frac{1-\Omega_{m0}}{1-(1-2\beta)\,e^\beta}\;,
\end{eqnarray}
where the limit $\beta=0$ corresponds to $\Lambda$CDM model, 
$\Omega_{m0}$ is  the dimensionless matter density parameter,
and $T_0 = T (z=0)$ is the current value for the torsion scalar.

By using  $T_0= -6 H_0^2$,  one finds from~(\ref{wec2}) the following WEC
constraint
\begin{eqnarray}\label{expmod}
\alpha\,\beta\, T_0\, e^\beta \geq 0\;.
\end{eqnarray}

Now we take $\Omega_{m0}=0.272^{+0.036}_{-0.034}$ ---which arises from the combination of
$557$ Type Ia Supernovae (SNe Ia) Union $2$ set, baryonic acoustic oscillation (BAO),
and the cosmic microwave background (CMB) radiation at $95\%$ confidence level
---along with the above observational value of  $H_0$. These values lead to
 $\beta>-1.256$ for the relation~(\ref{expmod}) to be satisfied.
Reciprocally, the inequality~(\ref{expmod}) is always fulfilled for all values $\beta$
such that $\beta>-1.256$. This makes explicit the constraint on parameter $\beta$ of the
exponential $f(T)$ gravity [Eq.(\ref{exp-grav})] for the WEC fulfillment.

\subsection{Born-Infeld  $f(T)$ gravity} \label{subsec42}
As the second concrete example, we consider the Born-Infeld (BI)
$f(T)$ gravity given by~\cite{biws}
\begin{equation}\label{largbi}
f(T)=\lambda\left[\left(1-\epsilon+\frac{2T}{\lambda}\right)^{1/2}-1\right]\;,
\end{equation}
where $\epsilon=4\Lambda/\lambda$ is a dimensionless parameter, $\Lambda$ is
the cosmological constant, and $\lambda$ is a
Born-Infeld-like constant. This gravity theory has been considered
in several cosmological contexts, which include the avoidance of
singularity in the standard model~\cite{avsig},
as a way to an inflationary scenario without inflaton~\cite{infla},
and also to bound the dynamics of the Hubble parameter~\cite{bisum}.
Clearly, the BI $f(T)$ gravity~(\ref{largbi}) reduces to the
standard TG (often referred to as TEGR) when $\lambda \rightarrow \infty $.
Here, we focus on the case $\lambda>0$~\cite{biws}. In this case, the WEC takes the form
\begin{equation}\label{biwec}
\epsilon-\frac{T_0}{\lambda}+\left(1-\epsilon+\frac{2T_0}{\lambda}\right)^{1/2}-1>0\;.
\end{equation}
This inequality holds for
\begin{equation} \label{BI-ranges}
0<\epsilon<1 \qquad \mbox{and} \qquad \lambda>-\frac{T_0}{\sqrt{\epsilon}(1-\sqrt{\epsilon})}\;,
\end{equation}
which makes apparent that the range of $\epsilon$ in which the WEC is fulfilled
coincides with that of an expanding universe where the cosmological constant is positive
(type $II$ of Ref.~\cite{biws}). Furthermore, by using  $T_0= -6 H_0^2$,  one finds
from inequations~(\ref{BI-ranges})  the WEC lower bound on the
parameter $\lambda$ in the BI teleparallel gravity, namely  $\lambda > 12.36\,$.

\section{Final remarks} \label{sec5}

Motivated by the attempts to explain the observed accelerating expansion of the
Universe with a modifying teleparallel gravitational theory, there have been many
recent papers on $f(T)$ gravity.
Despite the arbitrariness in the choice of different functional
forms of $f(T)$, which call for ways of constraining the possible  $f(T)$ gravity theories
on physical grounds, several features of $f(T)$ gravity have been discussed
in a number of recent articles.

In this paper we have proceeded further with the investigations on the potentialities,
difficulties, and limitations of $f(T)$ gravity theories by deriving  the classical
energy conditions in the $f(T)$ gravity context.
Starting from the Raychaudhuri equation along with the  requirement that gravity is
attractive, we have derived the null and strong energy conditions in the framework of
$f(T)$ gravity  and shown that, although similar, they differ from NEC and SEC
of general relativity, but in the limiting case $f(T)=T$, they reduce to well-known
NEC and SEC of Einstein's gravitational theory.
The comparison of the NEC and SEC inequalities [Eqs.~(\ref{sec1}) and (\ref{nec1})]
with those which would be obtained by translating these energy conditions in terms of
an effective energy-momentum tensor for $f(T)$ gravity, enabled us to obtain
the general expressions for the weak and dominant energy conditions
[Eqs.~(\ref{wec1}) and~(\ref{dec1})], which also reduce to the known corresponding
energy conditions in general relativity in the limit $f(T)=T$.

As concrete examples of how these energy conditions requirements may
constrain $f(T)$ gravity theories, we have discussed the WEC bounds on
two different $f(T)$ families of theories, namely the exponential and Born-Infeld
$f(T)$ gravity theories (Secs.~\ref{subsec41} and \ref{subsec42}). To this end,
we have used the current observational bounds on $H_0$ and  $\Omega_{m0}$
to show that the WEC are fulfilled for  $\beta>-1.256$ in the exponential
$f(T)$ gravity, whereas for Born-Infeld $f(T)$ gravity the 
WEC fulfillment is guaranteed for any $\lambda > 12.36\,$ such that
 $0<\epsilon<1$ holds.

Finally, we emphasize that although the energy conditions in $f(T)$ gravity
discussed in this paper have  well-motivated physical grounds (the attractive character
of gravity together with the Raychadhuri equation), the question as to
whether they should be employed to any solution of $f(T)$ gravity theories
is an open question, which  is ultimately related to the confrontation
between theory and observations. We recall that in the context of Einstein's
gravitational theory, this confrontation indicates
that all energy conditions seem to have been violated in the recent past
of cosmic evolution~\cite{EC-1,EC-8}.

\begin{acknowledgments}
M.J.R. acknowledges the support of FAPERJ under a CNE E-26/101.556/2010 grant.
This work was also supported by Conselho Nacional de Desenvolvimento
Cient\'{\i}fico e Tecnol\'{o}gico (CNPq) - Brasil, under grant No. 475262/2010-7.
M.J.R. thanks  CNPq for the Grant under which this work
was carried out. We are grateful to A.F.F. Teixeira for reading the manuscript
and indicating some omissions and typos. D.L. is particularly grateful to Professor P.X. Wu
for his several helpful suggestions and long-time support of the author's research.
\end{acknowledgments}


\end{document}